\documentstyle[twoside,fleqn,espcrc2,epsfig]{article}

\def\approxlt{\mathrel{\hbox{\rlap{\lower.55ex \hbox {$\sim$}}
        \kern-.3em \raise.4ex \hbox{$<$}}}}
\def\approxgt{\mathrel{\hbox{\rlap{\lower.55ex \hbox {$\sim$}}
        \kern-.3em \raise.4ex \hbox{$>$}}}}

\newcommand{\AmS}{{\protect\the\textfont2
  A\kern-.1667em\lower.5ex\hbox{M}\kern-.125emS}}

\title{BeppoSAX observations of low-energy spectral features in AGN}

\author{Astrid Orr\thanks{The SAX satellite is a joint Italian and Dutch 
programme. AO acknowledges an ESA Fellowship.}$^{\rm a}$, 
A.N. Parmar\address{Astrophysics Division, Space Science 
Department of ESA, ESTEC, P.O. Box 299, 2200 AG Noordwijk, The 
Netherlands}, 
T. Yaqoob\address{NASA Goddard Space Flight Center, Lab. for High 
Energy Astrophysics\\ 
Greenbelt, MD 20771, U.S.A.}, 
M. Guainazzi\address{BeppoSAX Science Data Center, A.S.I, c/o Nuova 
Telespazio, I-00131 Roma, Italy}
}

\begin{document}

\begin{abstract}
The combination of the broad band coverage and moderate spectral resolution
of the LECS and MECS instruments on-board BeppoSAX allow the spectra
of AGN to be studied in unprecedented detail down to  0.1 keV. 
We describe the calibration and the performance of the LECS  and 
 report on observations of 
low-energy absorption features in the spectra of both a low (MCG-6-30-15) and
a high luminosity (3C 273) AGN. These features provide important 
diagnostics on the location and nature of the material
surrounding the AGN. A comparison of LECS and ASCA/SIS low energy 
performance is also presented in the case of 3C 273.

\end{abstract}

\maketitle

\section{Introduction}
In the last decade, X-ray observations have provided evidence
for complex absorption and emission features in AGN spectra, in particular
below $\sim$10 keV. Examples of this complexity are the K$\alpha$ Fe  
emission \cite{Poun90} at  $\sim$ 6.4 keV and the photo-electric absorption 
from both neutral and highly ionized (warm) material in the line of sight and 
associated with the active nucleus \cite{Reyn97,Geor97}.   
The Einstein Observatory  and, later, the ROSAT Position Sensitive Proportional 
Counter (PSPC) were the first to
provide strong evidence for warm absorbers \cite{Halp84,Nand92}. 
This absorption was initially detected by the ROSAT PSPC 
in the spectrum of the Seyfert galaxy 
MCG-6-30-15 at an energy of 0.8 keV, consistent with a blend of 
O~{\sc vii} and O~{\sc viii} K-absorption edges.

The Low Energy Concentrator Spectrometer (LECS) is one of the narrow field
instruments (NFI) on-board BeppoSAX \cite{Parm97}. 
It operates in the energy range 0.1--10 keV. 
The LECS achieves this extended low-energy response by utilizing a driftless gas 
cell and an ultra-thin 1.25 $\mu$m entrance window. The LECS has particularly
good spectral resolution at energies $\approxlt$0.5~keV, where instruments 
such as the Solid State Imaging Spectrometer (SIS) on 
ASCA are not sensitive \cite{Tana94} and where instruments such as the 
ROSAT PSPC \cite{Trum83} (0.1--2.5~keV) have only moderate spectral 
resolution. The energy resolution of the LECS is 32\% 
(full width at half maximum, FWHM)
at 0.28~keV and 8.8\% at 6~keV. The angular resolution at these respective 
energies is $9.7'$ and $2.1'$.

The combination of  LECS data with simultaneous data from the other NFI on 
BeppoSAX, in particular the Medium Energy Concentrator Spectrometer 
\cite{Boel97} (MECS),
which has a higher effective area than the LECS above 1.8 keV and 
operates in the range 1.3--10 keV, provides a powerful tool for soft X-ray
spectroscopy of AGN.

\section{Calibration and performance of the LECS}

\begin{figure*}[htb]
\centerline{\hbox{\epsfig{figure=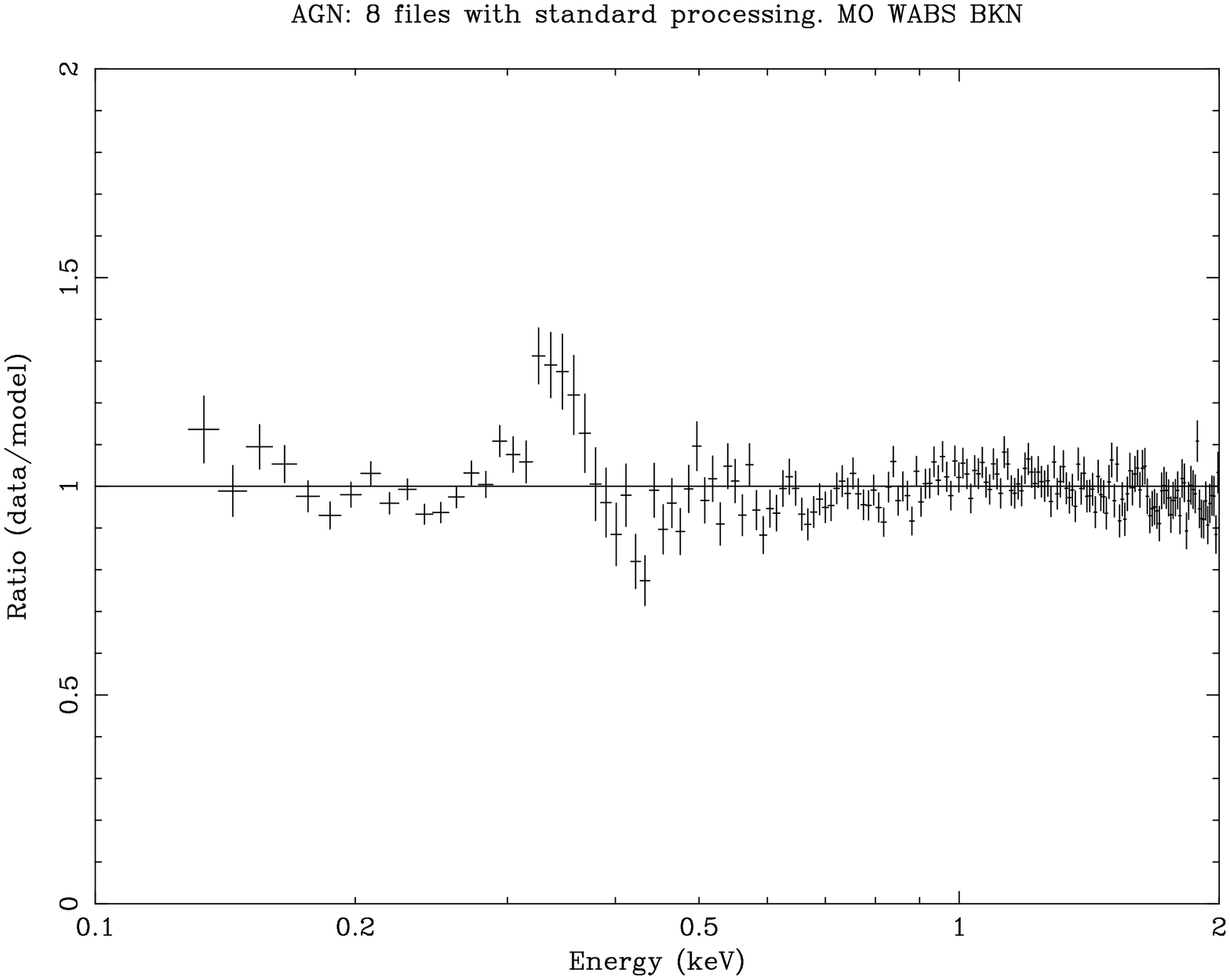,
height=8.5cm,width=6.5cm,angle=-90}}\hspace{0.1cm}
{\epsfig{figure=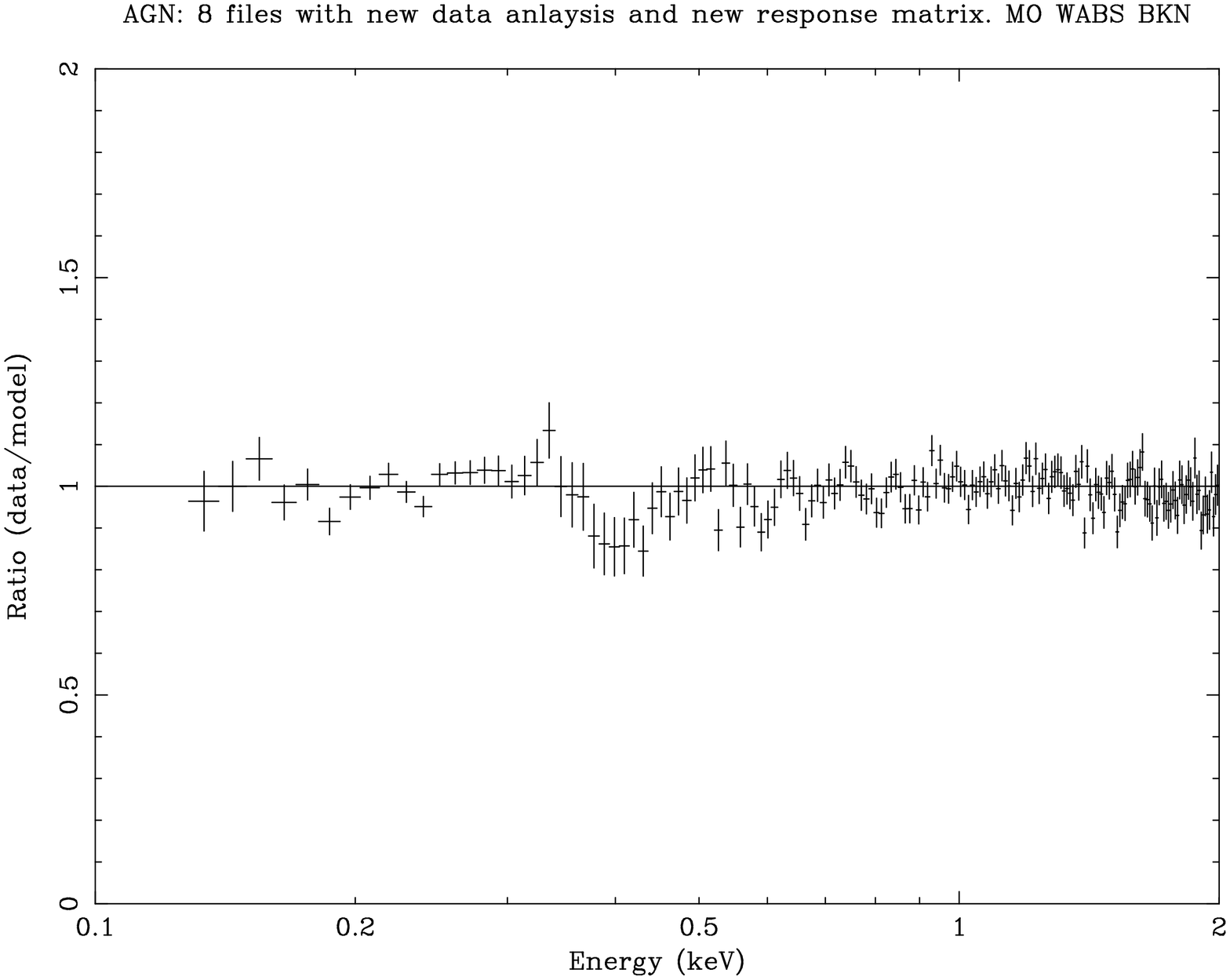,height=8.5cm,width=6.5cm,angle=-90}}}
\hspace{0.7cm}\centerline{\parbox{7.5cm}{Figure 1: Weighted average data-to-model ratio for a sample of 8 AGN. The spectra were obtained with pre-September 1997 processing and show a systematic feature between 0.3--0.5 keV.}
\hspace{1cm}\parbox{7.5cm}{Figure 2: Sample of 8 AGN. The spectra were 
obtained with the September 1997 release of LECS software and response 
matrix.}}
\label{fig:feature}
\end{figure*}

Great care has been taken in the ground calibration of the LECS
 \cite{Parm97}. 
However, it is difficult to verify the low-energy response of the LECS 
in-flight.
For instance the Crab Nebula is not detected by the LECS below 0.5~keV 
due to interstellar absorption. 
After more than a year of BeppoSAX operations sufficient data were 
acquired to enable a detailed study of the low energy calibration of the LECS
 \cite{Orra97}. 

Independently, much effort was spent on improving the LECS response matrix 
and LECS data analysis pipeline for the second release of the BeppoSAX/LECS  
software\footnote{Version V1.7.0 of SAXLEDAS of 01 September 1997.\\ 
See {\verb!ftp://astro.estec.esa.nl/pub/SAX/SOFTWARE!}.} 
and calibration\footnote{LECS response matrix release of 01 September
1997 (LEMAT V3.4.0). See response matrices in 
{\verb!ftp://astro.estec.esa.nl/pub/SAX/RESPONSE!}} in September 1997.

The changes to the LECS response matrix include an adjustment in the modeling 
of the LECS energy resolution to account for alterations in instrumental 
performance between pre-launch ground calibrations and in-orbit operations 
in a near-vacuum environment.
Improvements to the LECS data analysis pipeline involve a more detailed 
handling of the LECS burst-length correction \cite{Parm97,Lamm97}. 

Figure 1 shows the weighted average data-to-model ratio of a sample of 8
AGN whose BeppoSAX raw data (Final Observation Tapes) were readily available.
The data were prepared with the old, pre-September 1997 processing
(as available on the BeppoSAX data archive) and fit with
a broken power-law model including galactic and excess photo-electric 
absorption \cite{Orra97}.
A systematic feature is apparent at low energies, in fact, just where the
effective area of the LECS has a minimum value (see Fig. \ref{effarea}).
The maximum amplitude of fit 
residuals due to this effect is $\pm 25$\% of the folded model at a given 
energy between 0.3--0.5 keV. The feature is also very clearly seen in a 
larger sample of 26 blazar spectra, 
thereby confirming that its origin is instrumental. 

Figure 2 shows that the instrumental feature is strongly reduced using
the September 1997 release of the LECS software and response matrix.
The maximum amplitude is now 15\% between 0.4--0.5 keV. 
The remaining feature may be caused by uncertainties in modeling the LECS 
response at higher 
energies ($\approxgt$0.7 keV) where the effective area is much higher, or
by an underestimate of the amount of nitrogen in the LECS window.
Such effects will be included in forthcoming versions of the LECS response
matrix. In addition, we emphasize that there is no
systematic instrumental effect of amplitude $>$10\% in the energy range 
0.1--0.3 keV and above 0.5 keV.
\setcounter{figure}{2}
\begin{figure}[htb]
\centerline{\hbox{\epsfig{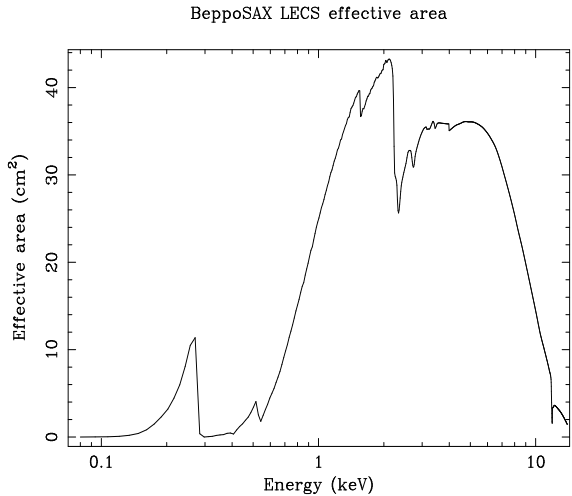}}\hspace{0cm}}
\caption{BeppoSAX LECS on-axis effective area as a function of energy.}
\label{effarea}
\end{figure}

\section{LECS/MECS observations of the warm absorber in MCG-6-30-15}

\begin{figure}[htb]
\vspace{-0.7cm}
\centerline{\hbox{\epsfig{figure=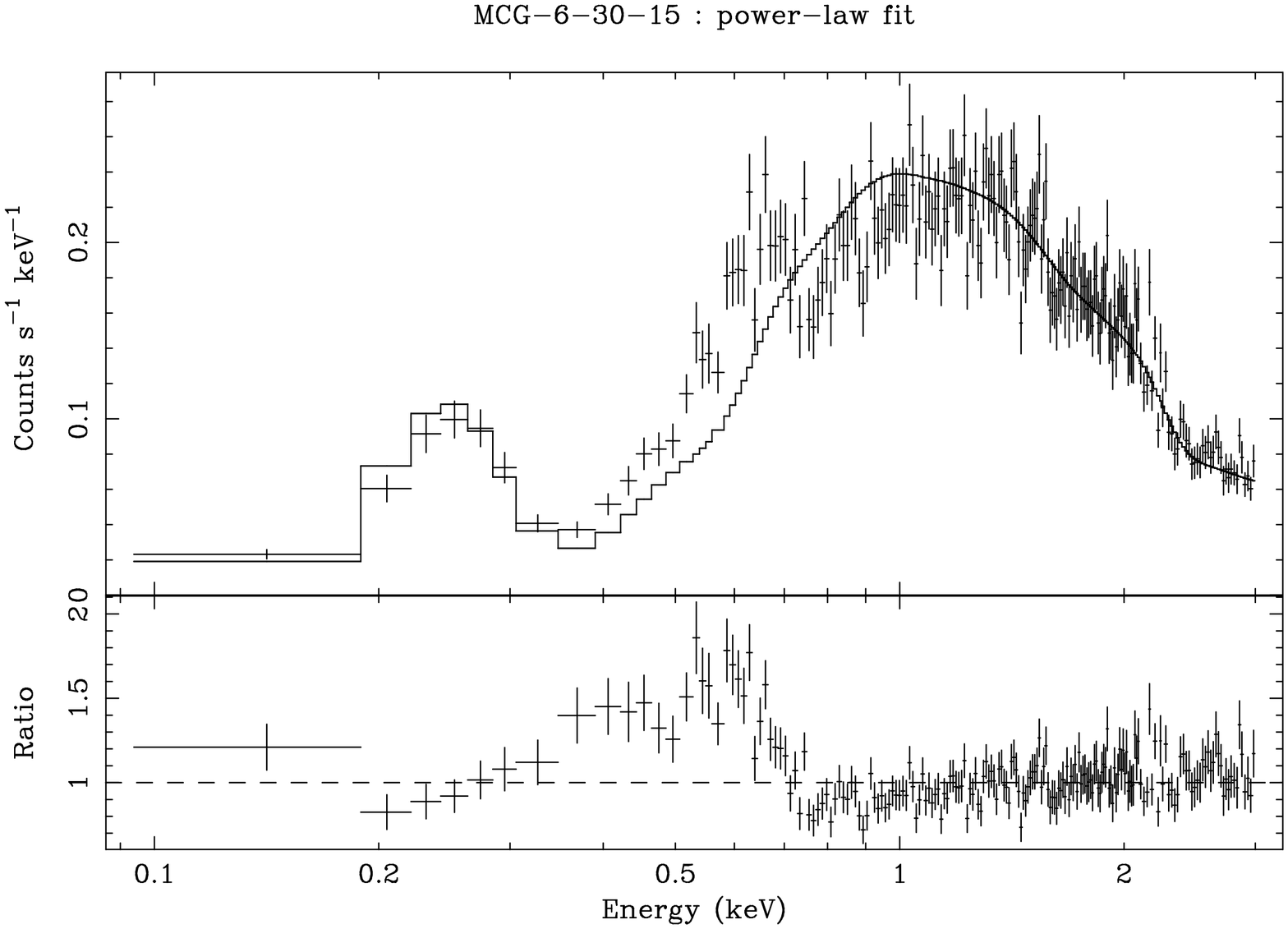,
height=8.5cm,width=7.5cm,angle=-90}}\hspace{0cm}}
\caption{Absorbed power-law fit to the LECS spectrum of MCG-6-30-15. 
Large residuals due to
a blend of O {\sc vii} and O {\sc viii} K-edges are clearly visible below 
1 keV.}
\label{mcgpow}
\end{figure}

\begin{figure}[htb]
\centerline{\hbox{\epsfig{figure=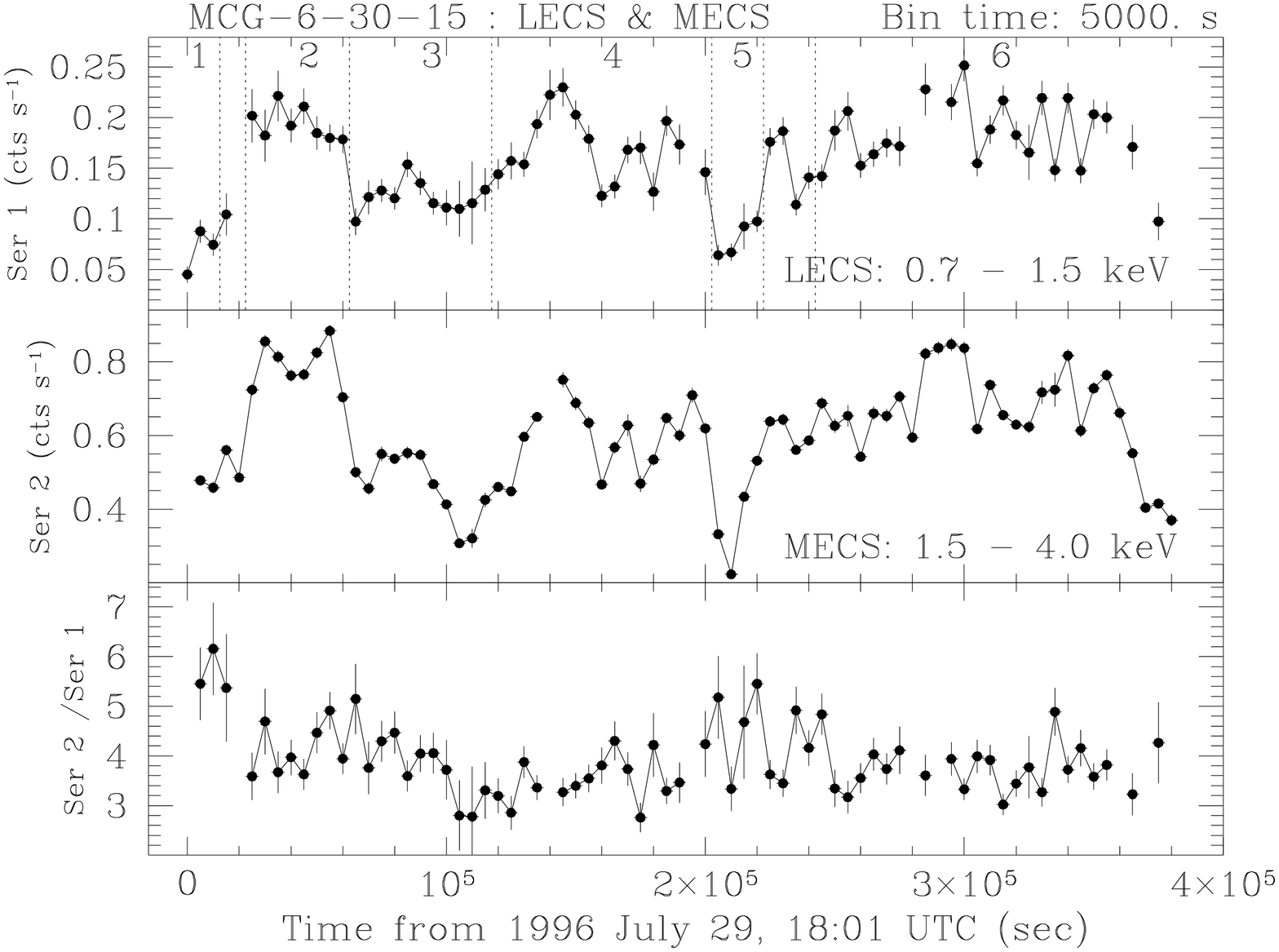,
height=8.5cm,width=6.5cm,angle=-90}}\hspace{0cm}}
\caption{Top: 0.7--1.5 keV LECS light curve of MCG-6-30-15. The epochs chosen
for a study of spectral variability are numbered from 1 to 6.
Middle: 1.5--4 keV MECS light curve. Lower: ratio of 1.5--4 keV counts over
0.7--1.5 keV counts.}
\label{lightc}
\end{figure}

\begin{figure}[htb]
\centerline{\hbox{\epsfig{figure=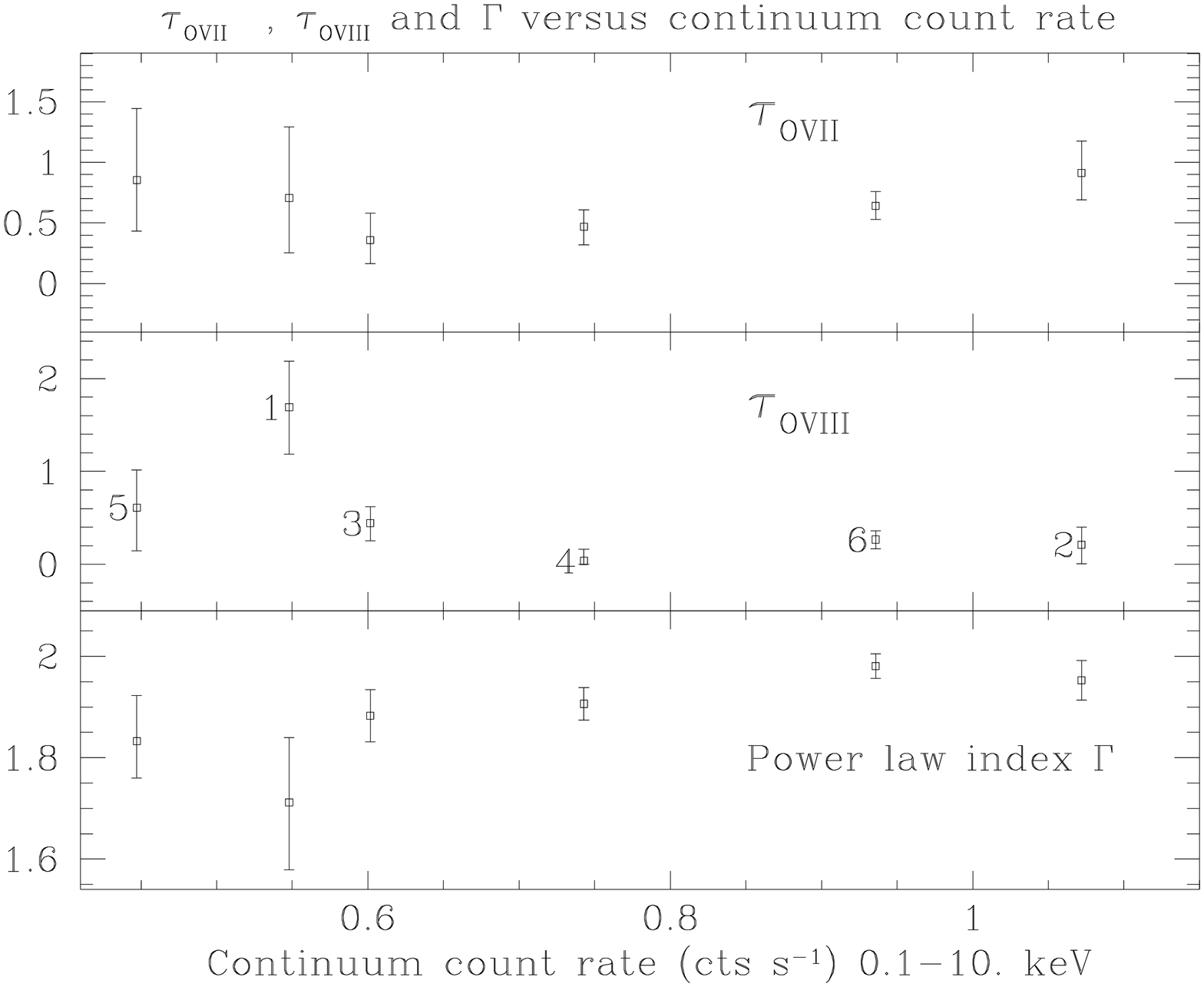,
height=8.5cm,width=6.5cm,angle=-90}}\hspace{0cm}}
\caption{MCG-6-30-15: plot of photon index $\Gamma$, 
$\tau_{{\rm O} {\sc vii}}$, $\tau_{{\rm O} {\sc viii}}$ 
as a function of un-absorbed model continuum 
count rate from 0.1--10 keV. Uncertainties are $\Delta \chi^2 = 1.0$.}
\label{tau}
\end{figure}

The Seyfert 1 galaxy MCG-6-30-15 was observed during the Science 
Verification Phase (SVP) of BeppoSAX between 1996 July 29 and August 3
\cite{Orrb97,Mole00}.
An analysis of the LECS and MECS data \cite{Orrb97} confirmed the presence of
a complex warm absorber. Figure \ref{mcgpow} shows that a 
simple power-law model with neutral absorption gives a poor fit to the spectrum
below 2 keV.  
The time averaged spectrum between 0.1--4 keV 
can be well described by a model composed of a power-law 
($\Gamma = 2.36\pm 0.38$), three absorption edges ($\chi^2= 398.1$ 
for 394 degrees of freedom, the uncertainties are $\Delta \chi^2 = 2.71$). 
Two edges have their energies fixed 
at the physical rest frame energies of the O~{\sc vii} and O~{\sc viii} 
K-edges (E$_1$ = 0.74 keV, E$_2$ = 0.87 keV, $\tau_1 = 0.91 \pm 0.18$, 
$\tau_2 = 0.03 \pm 0.15$) and the third edge is at E$_3 = 1.12 \pm 0.10$
with $\tau_3 = 0.19 \pm 0.08$ \cite{Orrb97}. 
The third edge is compatible with Ne {\sc ix} absorption at 1.20 keV.

The 0.7--1.5 keV and 1.5--4.0 keV light curves of MCG-6-30-15 are shown in 
Fig. \ref{lightc}, together with the corresponding hardness ratio.
Flux variations of a factor of 4 occurred during the observation. In fact,
in the first 7 hours of the observation the 0.7--1.5 keV count rate increased 
four-fold. The results of time resolved spectral analysis are shown in 
Fig. \ref{tau}. The main characteristics of the temporal behavior during
the BeppoSAX observation of MCG-6-30-15 can be summarized as following:
\begin{itemize}\vspace{-0.1cm}
\item{} The continuum flux and slope as well as the warm absorber are variable
 on time scales shorter than $2 \times 10^4$ s. \vspace{-0.2cm}
\item{} The variations in these two components are not always associated.\vspace{-0.2cm}

\item{} The optical depth of O~{\sc vii} remains approximately constant
throughout the observation.\vspace{-0.2cm}
\item{} The optical depth of O~{\sc viii} displays a significant change
during the low state at the start of the observation. Thereafter it remains
constant, even during  changes of continuum flux. \vspace{-0.2cm}
\end{itemize}

The dissimilar variability patterns of O~{\sc vii} and O~{\sc viii} have been
observed previously by ASCA \cite{Reya95,Otan96}. Multi-zone or stratified 
warm absorbers in photo-ionization equilibrium have been suggested as a 
possible explanation \cite{Reya95}.
The apparent lack of correlation of the optical depth of O~{\sc viii}
with changes in the continuum flux is different from what ASCA observed 
\cite{Otan96} and may indicate that simple photo-ionization
equilibrium does not apply. This, for instance, can be the case if the 
electron density of the warm absorber medium is low \cite{Reyb95}, thereby 
causing the recombination time-scale of the gas to be  longer than the
variability time-scale of the ionizing continuum source.  

\section{LECS/ASCA-SIS observations of the warm absorber in 3C 273}

\begin{figure}[htb]
\centerline{\hbox{\epsfig{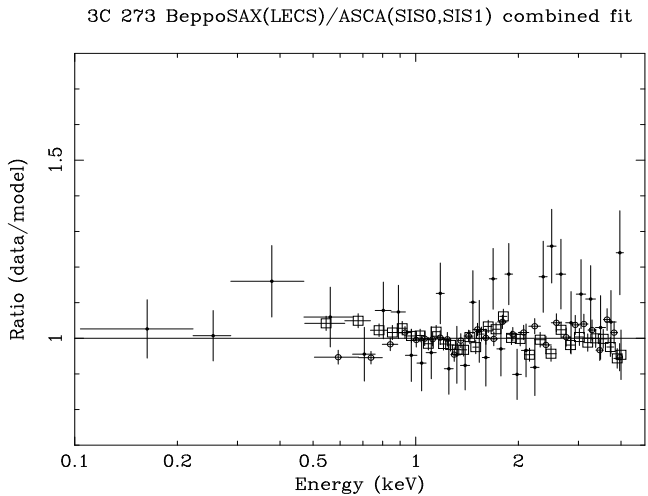}}\hspace{0cm}}
\caption{Data-to-model ratio of best fit of the ASCA-BeppoSAX X-ray spectrum
of 3C 273:
SIS0 (open squares) and SIS1 (open circles) between 0.5--4 keV, and LECS
(dots) data between 0.1--4 keV. See text.}
\label{3cfit}
\end{figure}

The bright quasar 3C 273 was observed quasi-simultaneously by
BeppoSAX and ASCA as part of dedicated program to cross-calibrate
the detectors on a number of different X-ray astronomy missions.
3C 273 was observed by BeppoSAX between 1996 July 18-21, 
during the SVP and by ASCA between 
1996 July 16-18. Simultaneous fits to the ASCA/SIS0/SIS1 and the BeppoSAX LECS
data show a very good agreement \cite{aorr00} between the data sets below 
4~keV.
In particular there is no evidence for any systematic deviation in the common
energy range 0.6--4 keV.
The LECS data above 4 keV have been ignored due to systematic effects 
caused by the off-axis LECS exposure.
Figure \ref{3cfit} shows the data-to-model ratio for the best-fit model
below 4 keV ($\chi^2$ = 491.1, with 424 degrees of freedom). 
The model is composed of two power-laws
($\Gamma_{soft} = 6.5 \pm 1.7$; $\Gamma_{hard} = 1.66\pm 0.02$), 
neutral absorption (galactic and excess) and an absorption edge
(E$_{edge} = 0.58\pm 0.02$ keV, in the source rest-frame; 
$\tau = 0.43 \pm 0.19$). 
The absorption at $\sim$0.6 keV in the present LECS data set is the first
unambiguous detection in 3C 273 of such a feature \cite{Gran97}.  
The LECS is the only instrument on BeppoSAX capable of detecting this
feature, and the SIS only detects the high energy ($\approxgt$0.6 keV) 
spectrum of the absorption. Therefore, the combination of SIS and 
LECS data provides a very good means of measuring the feature's spectral shape.

No other absorption edges are apparent in the range 0.1-4 keV. 
When attempting
to fit an absorption edge to a slight dip at $\sim$1.3 keV (source rest 
frame) in the SIS spectra in Fig. \ref{3cfit}, the improvement in fit 
statistics is insignificant
(less than 95~\% significance with the F-statistic) and the upper limit of
the optical depth is $\tau = 0.07$ (90\% confidence for one parameter of 
interest). 

A satisfactory identification of the single spectral feature must describe 
the spectrum self-consistently.
The feature may be identified with an O {\sc ii} or
an O {\sc iii} K-edge, which have physical rest frame energies of 0.58 and
0.6 keV, respectively. However, if the edge is O {\sc ii} the opacity from 
He {\sc i} may be significant (see e.g. discussion on absorber in PKS 2155-304,
based on the XSTAR phot-ionization code, \cite{Made00}). 
Finally, an energy shift due to bulk motion of the 
absorber (and therefore another identification) cannot be excluded.

\begin{thebibliography}{9}
\bibitem{Poun90} K. Pounds, et al., Nature 344 (1990) 132.
\bibitem{Reyn97} C. Reynolds, A. Fabian, MNRAS 273 (1995) 1167.
\bibitem{Geor97} I. George, et al., ApJS 114 (1997) 73. 
\bibitem{Halp84} J. Halpern, ApJ 281 (1984) 90. 
\bibitem{Nand92} K. Nandra, K. Pounds, Nature 359 (1992) 215.
\bibitem{Parm97} A.N. Parmar, et al., A\&AS 122 (1997) 309.
\bibitem{Tana94} Y. Tanaka, H. Inoue, S.S. Holt, PASJ 46 (1994) L37.
\bibitem{Trum83} J. Tr\"umper, Adv. Space Res. 2 (1983) 241.
\bibitem{Boel97} G. Boella, et al., A\&AS. 122 
(1997) 327.
\bibitem{Orra97} A. Orr, et al., 
BeppoSAX Science Data Center technical report - TR 015, 1997.   
\bibitem{Lamm97} U. Lammers, The SAX LECS Data Analysis System, Internal 
Report ESTEC SAX/LEDA/0010, ESA, 1997. 
\bibitem{Orrb97} A. Orr, S. Molendi, et al., A\&A 324 (1997) L77.
\bibitem{Mole00} S. Molendi et al., in preparation.
\bibitem{Reya95} C. Reynolds et al., MNRAS 277 (1995) 901.
\bibitem{Reyb95} C. Reynolds, A. Fabian, MNRAS 273 (1995) 1167. 
\bibitem{Otan96} C. Otani, et al., PASJ 48 (1996) 211. 
\bibitem{aorr00} A. Orr, et al., in preparation. 
\bibitem{Gran97} P. Grandi, et al., A\&A 325 (1997) L17.
\bibitem{Made00} G. Madejski, et al., submitted to ApJ. 
\end{thebibliography}
\end{document}